\documentclass[10pt,a4paper]{article}
\usepackage[utf8]{inputenc}
\usepackage{tabularray} 
\UseTblrLibrary{booktabs} 
\usepackage{physics}
\usepackage{siunitx}
\usepackage{floatpag}
\floatpagestyle{empty} 
\usepackage{placeins} 
\usepackage{jcappub_modified}
\usepackage[open]{bookmark} 

\notoctrue 

\newcommand{\Msolar}{\ensuremath{M_\odot}}

\newcommand{\planck}{\mathrm{Pl}}
\newcommand{\MP}{\ensuremath{M_\planck}}

\newcommand{\PBH}{\text{PBH}}
\newcommand{\DM}{\text{DM}}
\newcommand{\tot}{\text{tot}}

\title{Primordial black hole numbers: standard formulas and charts}

\date{Version 1.0}

\author{Eemeli Tomberg}

\affiliation{Consortium for Fundamental Physics, Physics Department,\\Lancaster University, Lancaster LA1 4YB, United Kingdom.}

\emailAdd{e.tomberg@lancaster.ac.uk}

\abstract{This brief note presents standard computations of primordial black hole mass $M$ given perturbations of scale $k$, and their late-time abundance $\Omega_\PBH$ given their initial density fraction $\beta$. I recap the assumptions made in these computations and present formulas and reference charts useful for a working cosmologist.}

\begin{document}

\maketitle

\section{Introduction}
\label{sec:intro}

\begin{table}
\label{tab:symbols}
\begin{center}
\begin{tblr}{width=\textwidth, colspec={X[1,r] X[0.2,c] X[1,l]}}
 Quantity & Symbol  & Values and notes \\ \midrule
 PBH mass & $M$ & \\
 Temperature & $T$ & at PBH formation \\
 & $T_0$ & today, \SI{2.726}{K} \\
 Energy density & $\rho$ & at PBH formation \\
 Hubble parameter & $H$ & at PBH formation \\
 & $H_0$ & \SI{70}{km/s/Mpc}, \cite{Tully:2023bmr} today \\
 & $H_I$ & during inflation \\
 Effective degrees of freedom & $g_{*}$ & 106.75 (full Standard Model) \\
 Effective entropy d.o.f.s & $g_{*s}$ & 106.75 (full Standard Model) \\
 & $g_{*s0}$ & 3.909 (today, with neutrinos) \\
 Comoving wavenumber & $k$ &  \\
 & $k_*$ & at CMB pivot scale, \SI{0.05}{Mpc^{-1}} \\
 E-folds of inflationary expansion & $N$ & from the CMB pivot scale \\
 Scale factor & $a$ & \\
 Redshift & $z$ & \\
 Frequency of gravitational waves & $f_\text{GW}$ & (stochastic background)\\
 Reduced Planck mass & $\MP$ & \SI{2.4354}{eV} \\
 Solar mass & $\Msolar$ & \SI{1.989e33}{g} \\
 PBH collapse efficiency factor & $\gamma$ & 0.2 \cite{Carr:1975qj} \\
 Initial PBH density parameter & $\beta$ & \\
 PBH density parameter today & $\Omega_\PBH$ & \\
 DM density parameter today & $\Omega_\DM$ & 0.264 \cite{ParticleDataGroup:2022pth} \\
 PBH dark matter fraction & $f_\PBH$ &  \\
 CMB tensor-to-scalar ratio & $r$ & $<0.036$ \cite{BICEP:2021xfz}  \\
\end{tblr}
\end{center}
\caption{Symbols and numerical values. For the numerical values, see \cite{ParticleDataGroup:2022pth, Planck:2018vyg}.}
\end{table}

When strong primordial perturbations enter the Hubble radius in the early universe, they may collapse into primordial black holes (PBHs). The time of entry depends on the perturbation scale $k$ and determines the mass of the PBHs, $M$. It also determines how much the initial PBH density gets diluted as space expands, relating the initial PBH mass fraction $\beta$ to the late-time mass fraction $\Omega_\PBH$. The formulas relating $M$ to $k$ and $\Omega_\PBH$ to $M$ and $\beta$ are standard (see e.g. \cite{Ballesteros:2017fsr, Karam:2022nym}), but the results may vary by up to a few orders of magnitude depending on the assumptions made, complicating comparisons between studies.
In this note, I derive these formulas in detail and present them in various easy-to-use numerical forms. The note is meant as a reference for the results and a resource for comparing PBH computations made under different assumptions. For further information about PBHs and early universe cosmology, I refer the reader to \cite{Carr:2021bzv, Carr:2024nlv, Green:2024bam, Mukhanov:2005sc, Weinberg:2008zzc, Lyth:2009zz, HelsinkiCosmologyNotes, Baumann:2018muz}.

I work in natural units where $\hbar=k_B=c=1$. Table \ref{tab:symbols} presents the symbols used in the note, together with benchmark values used as-is or as reference scales in the equations and charts. Conversions in natural units were done with the Mathematica code available at \url{www.eemelitomberg.net/NaturalUnits}, introduced in \cite{Tomberg:2021ajh}.

\section{PBH mass}
\label{sec:mass}

Consider PBHs that form when the universe is radiation-dominated with temperature $T$, energy density $\rho$, and Hubble parameter $H$. The PBH mass $M$ equals the mass inside the Hubble radius, up to a correction factor $\gamma$.\footnote{In truth, the perturbation strength affects the PBH mass, yielding a distribution of masses around our reference scale $M$, see e.g. \cite{Musco:2012au}.} We have the relations
\begin{equation} \label{eq:T_rho_H_M}
\begin{aligned}
    \rho &= \frac{\pi^2}{30}g_*T^4 = 3H^2\MP^2 \, , \\
    M &= \gamma\frac{4\pi}{3}H^{-3}\rho
    = \frac{4\sqrt{3}\pi\gamma\MP^3}{\sqrt{\rho}}
    = \frac{4\pi\gamma\MP^2}{H}
    = \frac{12\sqrt{10}\gamma\MP^3}{\sqrt{g_*}T^2} \\
    &= \SI{9.46e49}{g} \times \gamma\qty(\frac{T}{\unit{eV}})^{-2}\qty(\frac{g_*}{106.75})^{-1/2} \, .
\end{aligned}
\end{equation}
These PBHs form from primordial fluctuations corresponding to some comoving scale $k$, which exited the Hubble radius during inflation when
 \begin{equation}
     k = aH_I \, ,
 \end{equation}
where $a$ is the scale factor and $H_I$ is the Hubble parameter during inflation. I will relate all scales to the CMB pivot scale $k_* = \SI{0.05}{Mpc^{-1}}$ and write
\begin{equation} \label{eq:N_k}
    N \equiv \ln k/k_* \, .
\end{equation}
If $H_I$ stays approximately constant during inflation, this is the number of e-folds of inflation between the Hubble exits of $k$ and $k_*$.

To relate $k$ to $M$, we need to figure out when the scale re-enters the Hubble radius after inflation, that is, when $k=aH$ again. To deal with $a$, I consider adiabatic expansion between PBH formation and today, giving
\begin{equation} \label{eq:adiabatic_expansion}
    g_{*s}T^3a^3 = g_{*s0}T_0^3a_0^3 \quad \Rightarrow \quad a = a_0\frac{g_{*s0}^{1/3}T_0}{g_{*s}^{1/3}T} \, .
\end{equation}
Below, I also use the redshift $z \equiv a_0/a - 1$.
Here `0' refers to the quantities today, and I set $a_0=1$. Using \eqref{eq:T_rho_H_M} with $H=k/a$, and also $k=f_\text{GW}$ (the frequency of the related gravitational wave (GW) background and other similar signals)\footnote{Due to the general ambiguity related to the length scales, I neglect factors of $2\pi$ between $k$ and $f$.}${}^{,}$\footnote{For observational limits and prospects for GWs, see e.g. \cite{Moore:2014lga}.} then gives
\begin{equation} \label{eq:M_k_N}
\begin{aligned}
    M &= \gamma \frac{2\sqrt{2}g_*^{1/2}g_{*s0}^{2/3}\pi^2 T_0^2 \MP}{3\sqrt{5}g_{*s}^{2/3}k^2} \\
    &= \num{5.59e15}\Msolar \times \gamma \qty(\frac{k}{\SI{0.05}{Mpc^{-1}}})^{-2}\qty(\frac{g_{*s}^4 g_*^{-3}}{106.75})^{-1/6} \\
    &= \SI{1.11e49}{g} \times \gamma \times e^{-2N}\qty(\frac{g_{*s}^4 g_*^{-3}}{106.75})^{-1/6} \\
    &= \num{1.32e-15}\Msolar \times \gamma \qty(\frac{f}{\unit{Hz}})^{-2}\qty(\frac{g_{*s}^4 g_*^{-3}}{106.75})^{-1/6} \, .
\end{aligned}
\end{equation}

\newpage

\section{PBH abundance}
\label{sec:abundance}
I assume the PBHs form instantaneously with mass $M$ when the scale $k$ re-enters the Hubble radius. I call their initial energy density fraction $\beta$, as is standard in the literature. The initial PBH energy density is then $\beta\times\rho$. Today, the energy density has diluted by factor $a^3$ and the total energy density of the universe is $3H_0^2\MP^2$. Today's PBH density parameter is then
\begin{equation} \label{eq:Omega_PBH}
\begin{aligned}
    \Omega_{\PBH,0} &= \frac{\beta\rho a^3}{3H_0^2\MP^2} \\
    &= \frac{2^{1/4}\pi^2\beta\gamma^{1/2}g_*^{3/4}g_{*s0}T_0^3}{3\sqrt{3}5^{3/4}g_{*s}H_0^2\sqrt{M \MP}} \\
    &=
    \num{9.16e7} \times \beta \gamma^{1/2} \qty(\frac{M}{\Msolar})^{-1/2}\qty(\frac{g_{*s}^4g_{*}^{-3}}{106.75})^{-1/4} \qty(\frac{H_0}{\SI{70}{km/s/Mpc}})^{-2} \\
    &=
    \num{4.09e14} \times \beta \gamma^{1/2} \qty(\frac{M}{\SI{e20}{g}})^{-1/2}\qty(\frac{g_{*s}^4g_{*}^{-3}}{106.75})^{-1/4} \qty(\frac{H_0}{\SI{70}{km/s/Mpc}})^{-2}\, ,
\end{aligned}
\end{equation}
where I used the results of the previous section.

For a wide PBH mass distribution, it is conventional to keep track of the contribution to $\Omega_\PBH$ from PBHs in a mass bin of logarithmic width $\dd \ln M$ and define
\begin{equation} \label{eq:f_PBH}
    f_\PBH(M) \equiv \frac{1}{\Omega_\DM}\frac{\dd \Omega_{\PBH,0}}{\dd \ln M} \, .
\end{equation}
The total PBH dark matter fraction is then
\begin{equation} \label{eq:f_PBH_tot}
    f_\PBH^\tot \equiv \frac{\Omega_{\PBH,0}}{\Omega_\DM} = \int \dd \ln M \, f_\PBH(M) \, .
\end{equation}
PBH abundance bounds are often given in terms of $f_\PBH$.

In the above computation, I anchored the early universe quantities to physical scales by using the cosmological observables today, in particular, $T_0$ and $H_0$. In the literature, the computation is sometimes anchored to the matter--radiation equality instead, see e.g. \cite{Kannike:2017bxn, Rasanen:2018fom}. Other anchorings are also possible. Instead of $T_0$ and $H_0$, one simply has to know the redshift (or $k$) and the energy densities of all the matter components at the reference scale.

The next page presents a chart of observational limits on $f_\PBH$ for various PBH masses, pulled from \cite{KavanaughPBHbounds}, assuming all the PBHs have the same mass. Below the chart, the mass is mapped to various other quantities from Table~\ref{tab:symbols} using the table's numerical values and the formulas above. In particular, $g_{*}=g_{*s}=106.75$ at PBH formation is used throughout for simplicity. The following page presents an extended version of the mapping.\footnote{At the low end of the temperature scale, matter becomes comparable to radiation, and the equations of this note break down. I neglect this problem since the corresponding PBHs are, in any case, unrealistically massive.}

\begin{figure}
    \centerline{\includegraphics[scale=1]{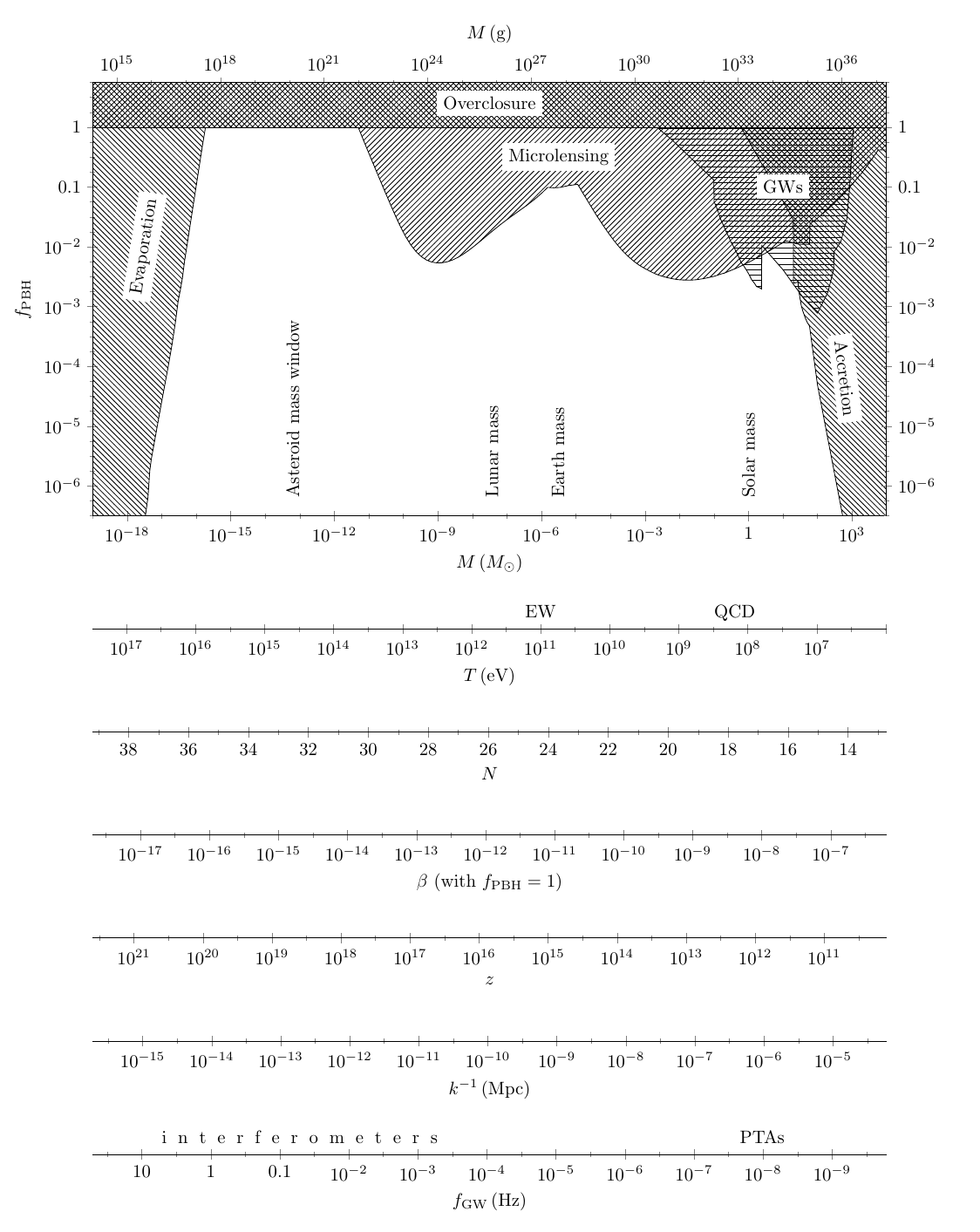}}
    \label{fig:fPBH}
\end{figure}

\begin{figure}
    \vspace{-2cm}
    \centerline{\includegraphics[angle=90, scale=1]{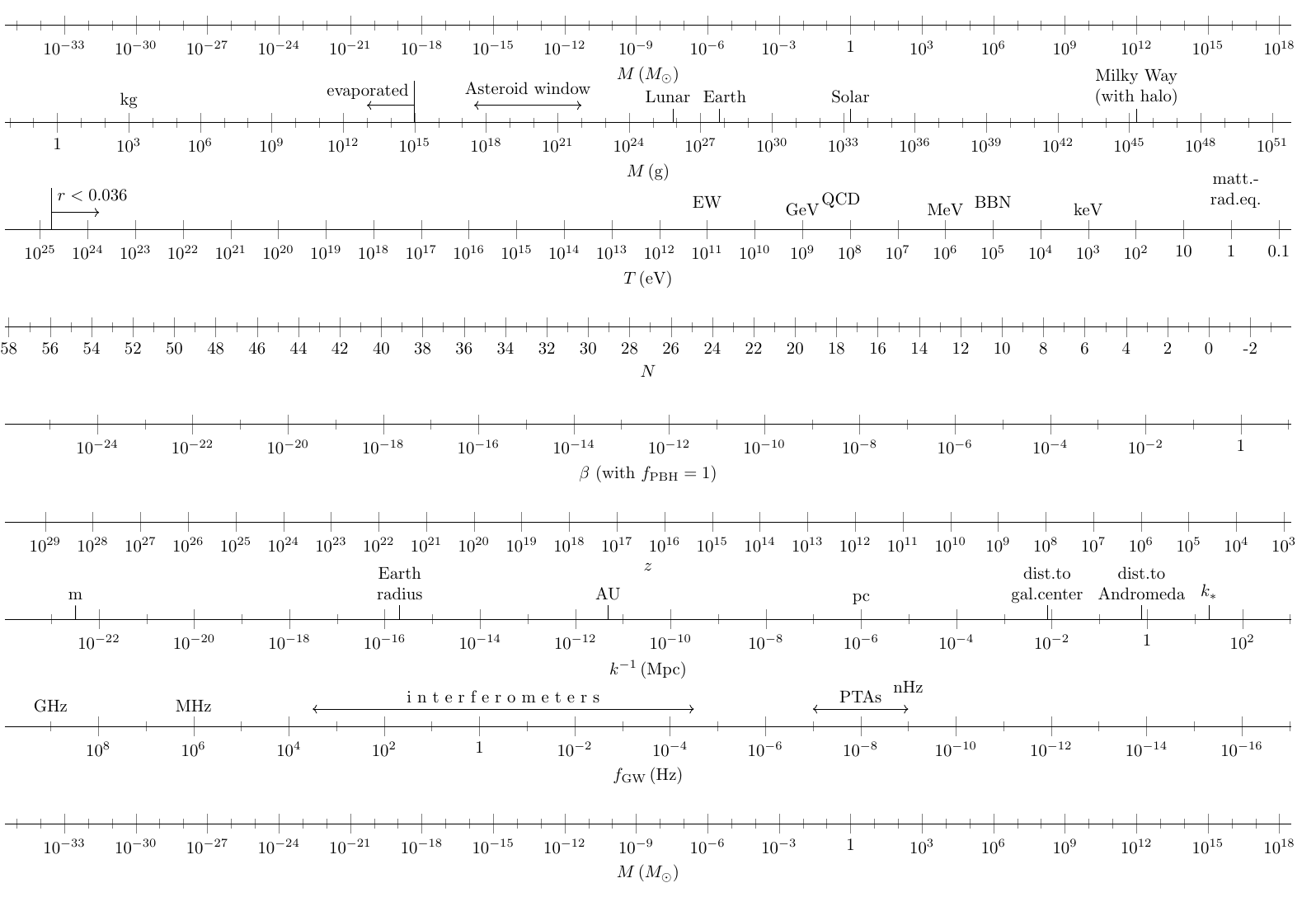}}
    \label{fig:scales}
\end{figure}

\section{Beyond the standard scenario}
In this note, I have dealt with three sets of linked variables: $M$, $\rho$, and $H$ linked by geometry and the Friedmann equations; $a$ and $z=a_0/a - 1$; and $N$, $f_\text{GW}$, and $k$ linked directly by \eqref{eq:N_k} and $f_\text{GW}=k$. The last are related to the others through $k=aH$, but the rest of the relations go through the temperature $T$, and they may be modified in non-standard cosmological scenarios. Relating $a$ to $T$ by \eqref{eq:adiabatic_expansion} requires a thermal bath undergoing adiabatic expansion, isolated from possible extra sectors (up to changes in $g_{*s}$). Further relating $T$ to $M$ through \eqref{eq:T_rho_H_M} requires the thermal bath to dominate over other sectors. Hidden sectors decoupled from the Standard Model can severe these relations, as can non-standard expansion history, such as an early period of matter or dark energy domination or kination. In such cases, equations \eqref{eq:M_k_N} and \eqref{eq:Omega_PBH} have to be reworked.

\FloatBarrier

\bibliographystyle{JHEP}
\bibliography{conv}

\end{document}